\documentclass[
aps,
prd,
reprint,
superscriptaddress,
longbibliography
]{revtex4-2}

\usepackage[utf8]{inputenc}
\usepackage[T1]{fontenc}
\usepackage{lmodern}
\usepackage[english]{babel}
\usepackage{graphicx}
\usepackage{amsmath,amssymb}
\usepackage{physics}
\usepackage{siunitx}
\AtBeginDocument{\RenewCommandCopy\qty\SI}
\usepackage{subcaption}
\usepackage[font=small,labelfont=bf,justification=justified,format=plain,singlelinecheck=false,compatibility=false]{caption}
\usepackage{booktabs}
\usepackage{microtype}
\usepackage[hidelinks]{hyperref}
\usepackage{makecell}
\usepackage{url}
\usepackage{ulem}
\usepackage{multirow}

\begin{document}

\preprint{preprint for arXiv}

\title{Measurement of Light Yield Response of Gd-compatible Water-based Liquid Scintillator with the Brookhaven 1-ton testbed}

\author{S. Gwon}
\affiliation{Chung-Ang University, Seoul, South Korea}


\author{M. Askins}
\affiliation{Physics Department, University of California at Berkeley, Berkeley, CA 94720-7300, USA}
\affiliation{Lawrence Berkeley National Laboratory, 1 Cyclotron Road, Berkeley, CA 94720-8153, USA}

\author{D.M. Asner}
\affiliation{Brookhaven National Laboratory, Upton, NY, USA}

\author{A. Baldoni}
\author{D.F. Cowen}
\affiliation{Physics Department, The Pennsylvania State University, State College, PA 16801, USA}

\author{R.~Diaz Prerez}
\affiliation{Brookhaven National Laboratory, Upton, NY, USA}

\author{M.V. Diwan}
\affiliation{Brookhaven National Laboratory, Upton, NY, USA}

\author{S. Gokhale}
\affiliation{Brookhaven National Laboratory, Upton, NY, USA}

\author{S. Hans}
\affiliation{Brookhaven National Laboratory, Upton, NY, USA}
\affiliation{Bronx Community College, Bronx, NY 10453, USA}


\author{P. Kumar}
\affiliation{Department of Physics and Astronomy, University of Alabama, Tuscaloosa, AL 35487, USA}

\author{G. Lawley}
\affiliation{Physics Department, Stony Brook University, Stony Brook, NY 11794, USA}

\author{S. Linden}
\affiliation{Brookhaven National Laboratory, Upton, NY, USA}

\author{G.D. Orebi Gann}
\affiliation{Physics Department, University of California at Berkeley, Berkeley, CA 94720-7300, USA}
\affiliation{Lawrence Berkeley National Laboratory, 1 Cyclotron Road, Berkeley, CA 94720-8153, USA}

\author{J. Park}
\affiliation{Chung-Ang University, Seoul, South Korea}

\author{C. Reyes}
\affiliation{Brookhaven National Laboratory, Upton, NY, USA}

\author{R. Rosero}
\affiliation{Brookhaven National Laboratory, Upton, NY, USA}


\author{K. Siyeon}
\affiliation{Chung-Ang University, Seoul, South Korea}

\author{M. Smiley}
\affiliation{Physics Department, University of California at Berkeley, Berkeley, CA 94720-7300, USA}
\affiliation{Lawrence Berkeley National Laboratory, 1 Cyclotron Road, Berkeley, CA 94720-8153, USA}



\author{J.J. Wang}
\affiliation{Department of Physics and Astronomy, University of Alabama, Tuscaloosa, AL 35487, USA}

\author{M. Wilking}
\affiliation{School of Physics and Astronomy, University of Minnesota, Minneapolis, MN  55455, USA}

\author{G. Yang}
\thanks{Corresponding author}
\affiliation{Brookhaven National Laboratory, Upton, NY, USA}

\author{M. Yeh}
\affiliation{Brookhaven National Laboratory, Upton, NY, USA}

\date{\today}

\begin{abstract}
The Water-based Liquid Scintillator (WbLS) enables hybrid detection by combining scintillation and Cherenkov signals, providing superior event reconstruction capabilities compared to conventional neutrino detectors.
We measured the light yield of Gd-compatible WbLS at varying concentrations from 0.35\% to 1\% by mass, using cosmic-ray muons in a 1-ton scale detector at BNL.
The light yield is measured as (69.16 $\pm$ 6.92) ph / MeV at 0.35\% concentration, which increased to (87.32 $\pm$ 8.73) ph / MeV at 1\%.
These results establish a quantitative basis for optimizing future WbLS-based detectors in neutrino physics.
\end{abstract}

\maketitle

\section{Introduction}
Large, kiloton-scale detectors are being planned and built around the world to investigate a wide range of physics, including the search for Beyond-Standard-Model phenomena, the detection of astrophysical neutrinos from sources like supernovae, and precision measurements of neutrino oscillations~\cite{hyperk,juno,DUNE_TDR}.
They aim to precisely measure neutrino oscillation, proton decay, as well as serve as active shielding for dark matter direct detection experiments.
The detector material that is widely used for such a large detector so far is a binary choice:
either water, leveraging directional Cherenkov radiation, or organic liquid scintillator, producing a high yield of isotropic light.
These technologies present a trade-off. 
While Water-Cherenkov detectors are easily scalable to huge volumes at low cost and provide excellent particle directionality, their high energy threshold and low light yield render them insensitive to low-energy events.
Liquid scintillators overcome this limitation with their excellent light yield and correspondingly low energy threshold, but the scalability of liquid scintillators is limited by the high cost, chemical hazards, and environmental concerns of organic materials. 
This fundamental dilemma—a choice between cost-effective scalability and low-energy sensitivity—motivates the development of a hybrid technology.

WbLS has been proposed as a next-generation detection medium to compensate for the shortcomings of both techniques while maximizing their advantages~\cite{chess, BUTTON, Theia:2019non, Eos:2022wbls}.
WbLS is a medium in which a liquid scintillator is dispersed throughout pure water in the form of micelles.
Based on this, it is expected to improve light yield and lower the energy threshold by scintillation, while retaining the existing infrastructure of conventional water-Cherenkov detectors.
\textcolor{black}{Moreover, the aqueous base allows efficient loading of neutron-capturing compounds such as gadolinium (Gd).
The identification of neutron-capture events improves background discrimination and event reconstruction in neutrino measurements.}
In addition, the WbLS concentration can be tuned with precision to meet specific experiments' goal.
While previous study characterized the light yield and stability of 1\% WbLS~\cite{2022_WbLS_paper}, this study provides the first measurement of light yield as a function of Gd-compatible WbLS concentration from 0.35\% to 1\% by mass. 
These results establish a benchmark for optimizing future large-scale WbLS-based detectors. A similar measurement was performed and showed a comparable result with the BNL 1-ton testbed measurement~\cite{Rosero2024_GdWbLS}.

\section{1-ton WbLS Detector}
\label{sec:1-ton_detector}
\begin{figure}[htbp]
    \centering
    \begin{subfigure}[b]{0.43\linewidth}
        \centering
        \includegraphics[width=\linewidth]{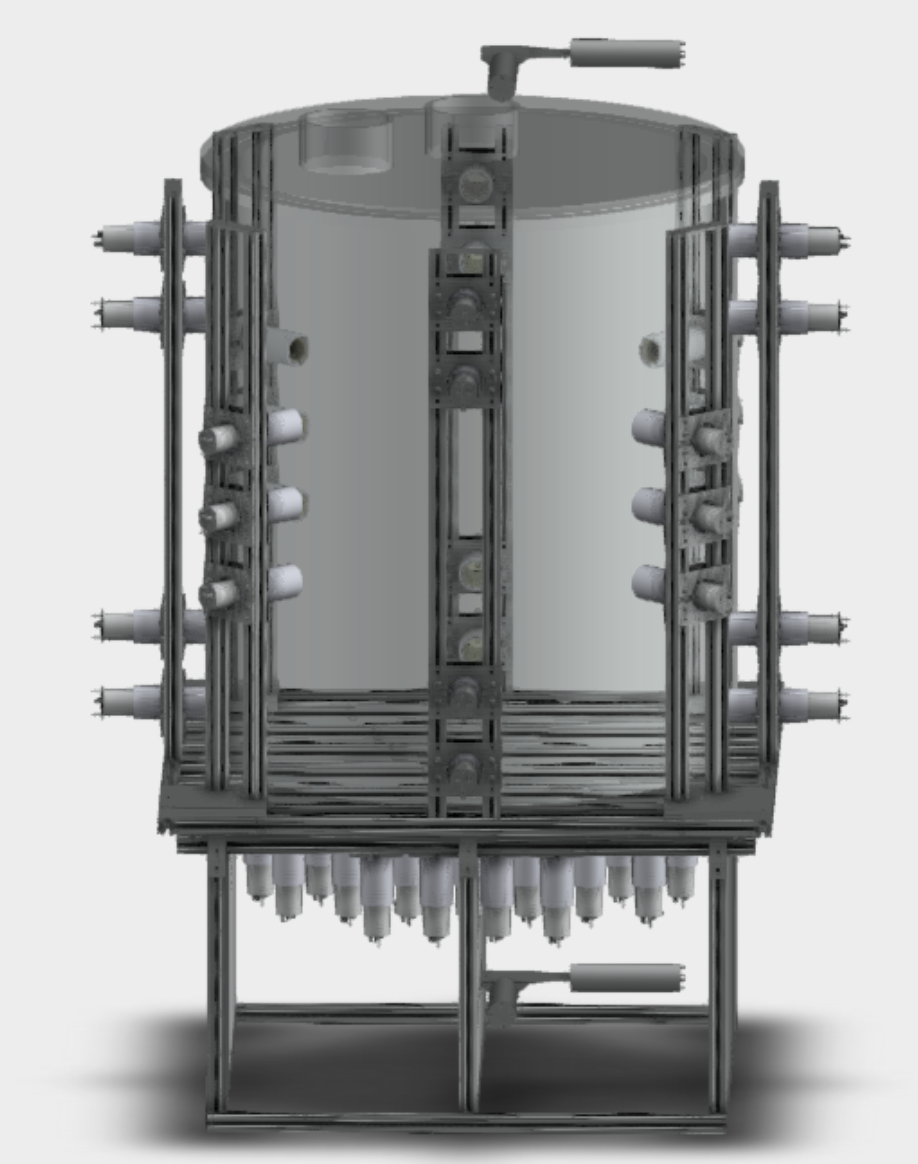}
        \caption{}
        \label{fig:1ton_schematic}
    \end{subfigure}
    \hfill
    \begin{subfigure}[b]{0.5\linewidth}
        \centering
        \includegraphics[width=\linewidth]{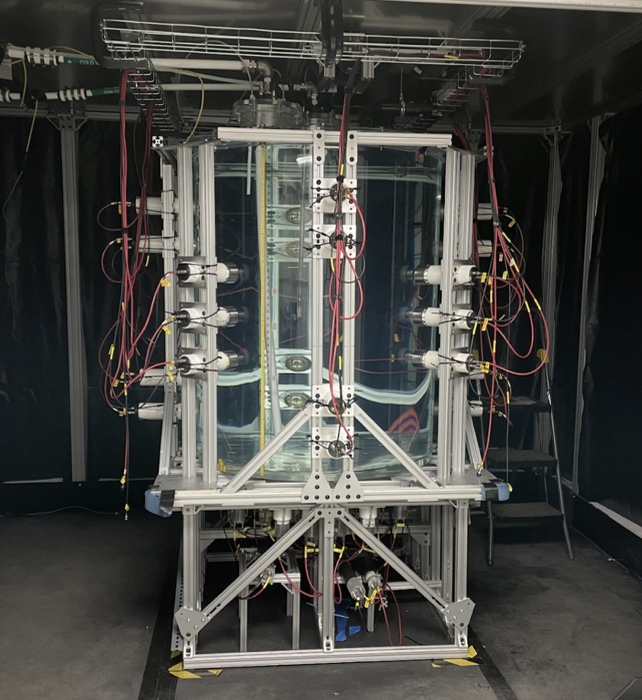}
        \caption{}
        \label{fig:1ton_photo}
    \end{subfigure}
    \caption{(a) Schematic of the 1-ton detector. (b) Photograph of the 1-ton detector installed inside a light-tight dark box.}
    \label{fig:detector_setup}
\end{figure}
The experiment was performed using the 1-ton WbLS demonstrator, built at Brookhaven National Laboratory (BNL) in 2022.
It consists of a cylindrical transparent acrylic (Nakano UVT) tank with an outer radius of 575 mm and a height of 1275 mm.
The tank is instrumented with a total of 58 photomultiplier tubes (PMTs), which are optically coupled to the acrylic surface with silicone gel pads referred to as ``cookies''.
Thirty 2-inch PMTs are placed on the bottom, and twenty-eight 3-inch PMTs (model 9821B) are arranged in seven rows on the side (two upper, three central, and two lower rows) as shown in Fig.\ref{fig:detector_setup}.

\begin{figure}[htbp]
    \centering
    \includegraphics[width=1.0\linewidth]{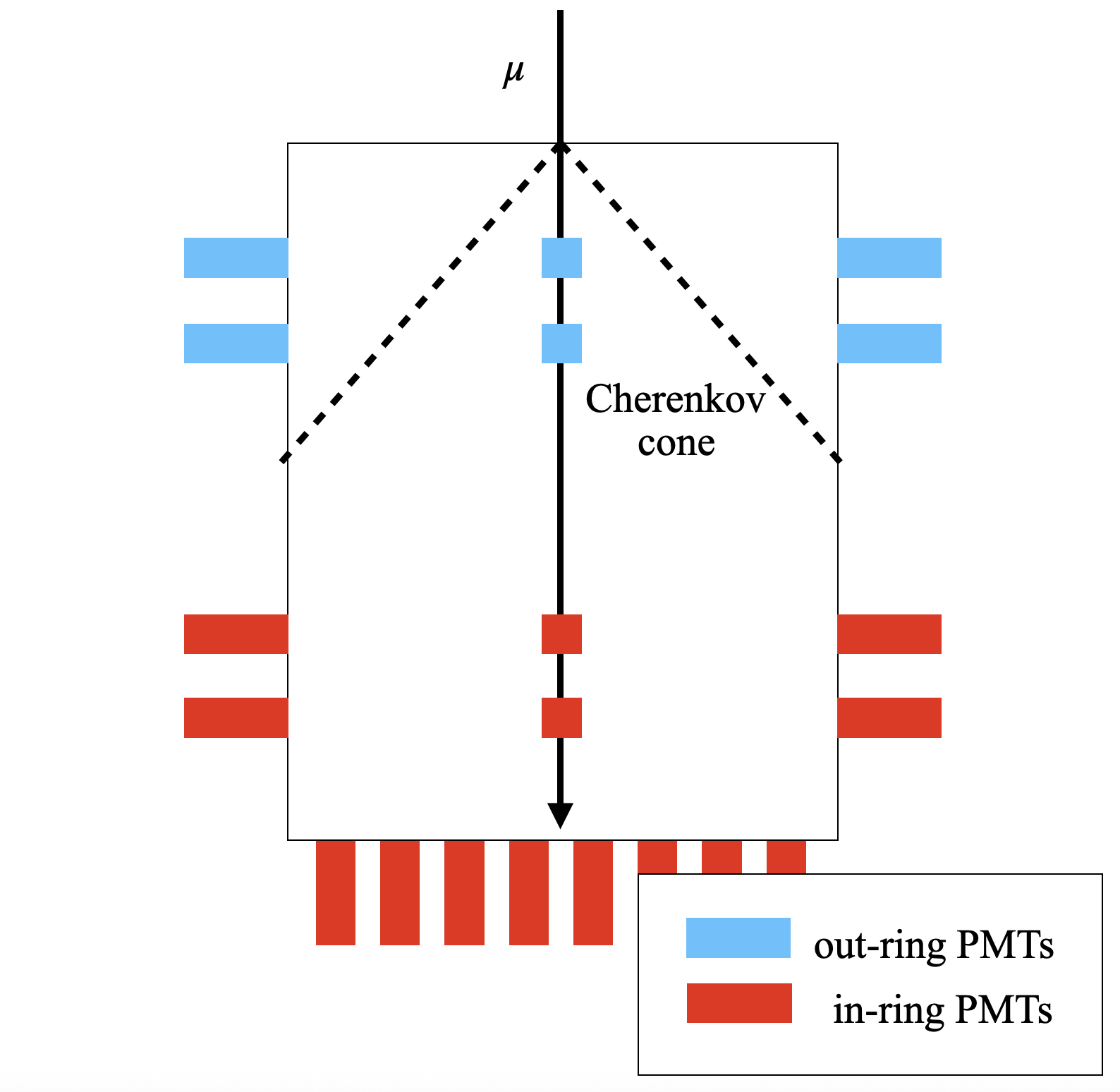}
    \caption{Conceptual diagram of the 1-ton WbLS detector showing the distinction between in-ring and out-ring PMT regions. The dashed lines represent the Cherenkov cone emitted by a through going muon, and the solid arrow indicates the muon's path. Out-ring PMTs (blue) primarily detect isotropically emitted non-Cherenkov light (mostly scintillation light), whereas in-ring PMTs (red) are sensitive to both Cherenkov and non-Cherenkov light.}
    \label{fig:inring_outring}
\end{figure}

Due to this unique geometry, Cherenkov light from vertically through-going muons (reffered to ``crossing muon'') from top to bottom mainly reaches the bottom and side two rows of PMTs, while the top two rows are dominated by scintillation light. 
Based on this distribution, the PMTs are classified into two groups: those inside the Cherenkov cone of the crossing muon (“in-ring”) and those outside the cone (“out-ring”) as shown in Fig.~\ref{fig:inring_outring}.
This classification enables the disentanglement of the scintillation and Cherenkov components, which can subsequently be utilized for light yield analysis.

To ensure stable optical properties of the WbLS for the duration of the measurement, a circulation system was used to maintain liquid purity, and the internal temperature was continuously monitored.
Further details of the experimental setup can be found in the previous publication~\cite{2022_WbLS_paper}.

\begin{figure}
    \centering
    \begin{subfigure}[b]{0.45\linewidth}
        \centering
        \includegraphics[width=\linewidth]{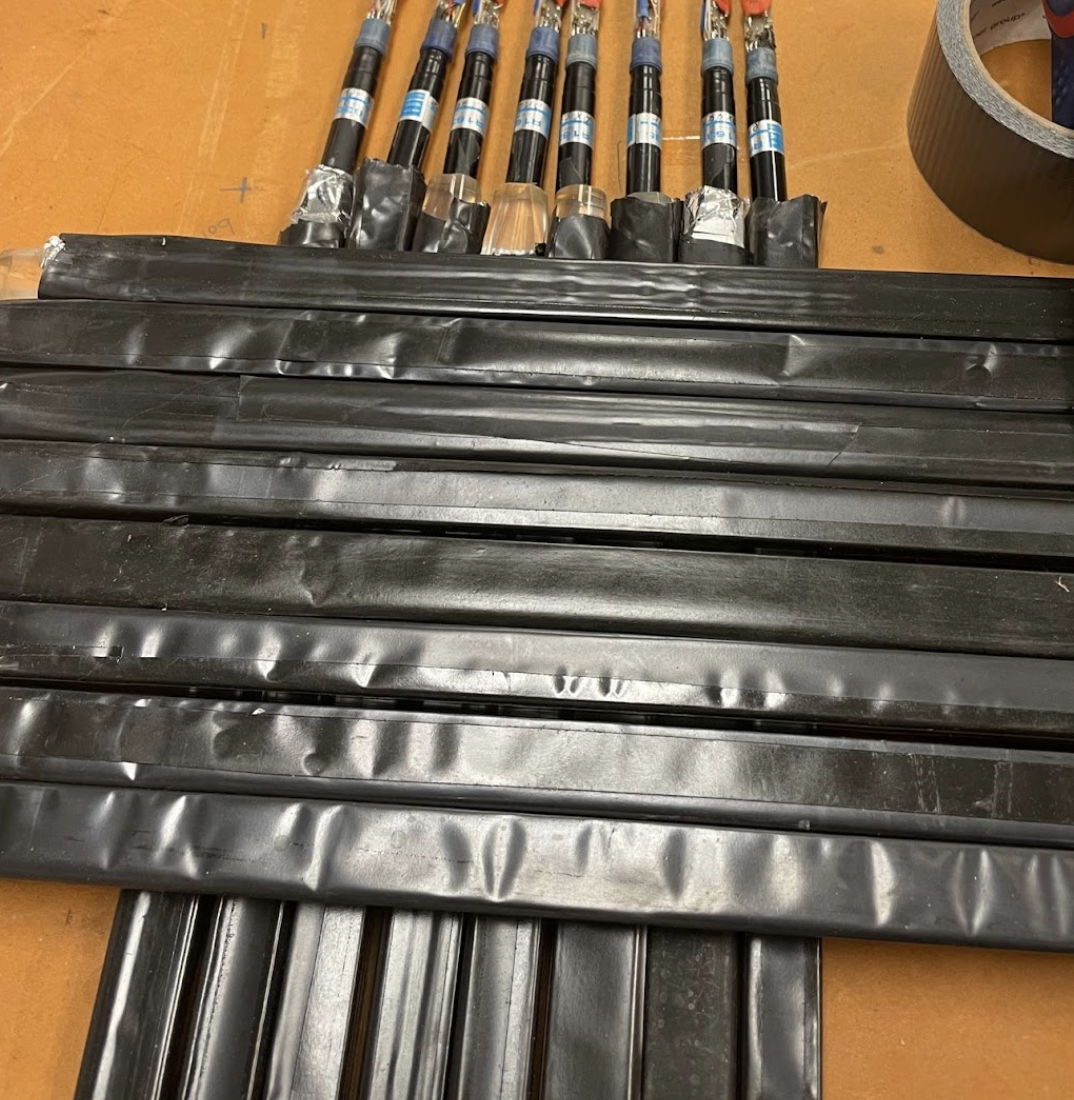}
        \caption{}
        \label{fig:hodoscope_1}
    \end{subfigure}
    \hfill
    \begin{subfigure}[b]{0.50\linewidth}
        \centering
        \includegraphics[width=\linewidth]{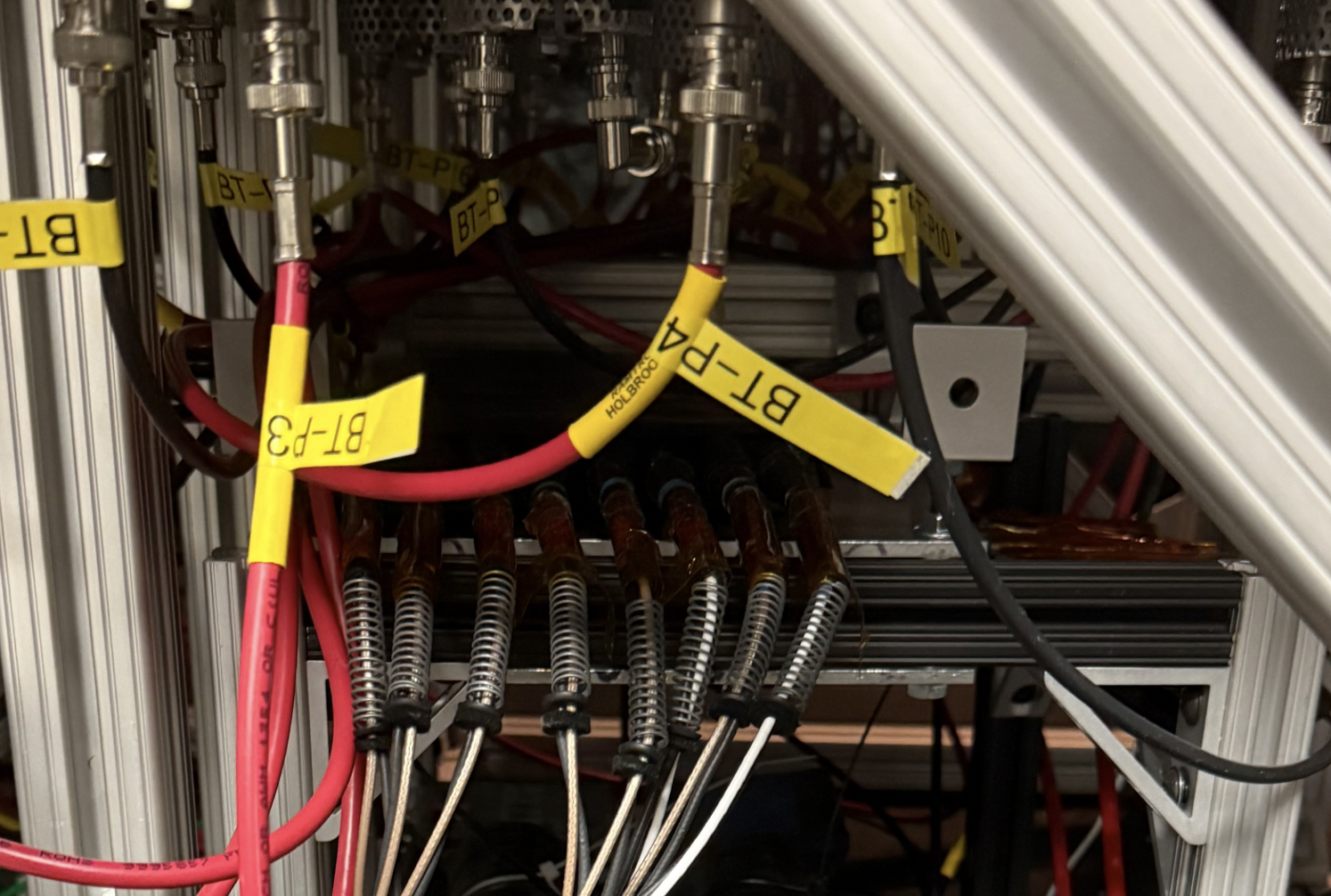}
        \caption{}
        \label{fig:hodoscope}
    \end{subfigure}
    \caption{(a) Two orthogonal layers of hodoscope array. (b) The bottom hodoscope array, positioned below the tank.}
    \label{fig:hodoscope}
\end{figure}
To improve muon track reconstruction, additional segmented plastic scintillator arrays, serving as a hodoscope, were installed at both the top and bottom of the tank. 
Each hodoscope consists of two layers with eight 1-inch-wide scintillator bars.
one layer is consists of eight plastic scintillator bars.
Two layers oriented orthogonally, creating a 15 cm $\times$ 15 cm overlapping region with 8 $\times$ 8 segmentation, providing higher spatial resolution.
Fig.~\ref{fig:hodoscope} shows the bottom hodoscope installed beneath the PMTs. 
The hodoscopes aim to reconstruct the precise entry and exit location of muons, which reduces muon angle uncertainty significantly.
However, the strict event selection required by the hodoscopes leads to lower statistics. Therefore, this hodoscope-tagged data is not used in the main analysis but serves as a independent set for validation.

\section{Data Acquisition}
\label{sec:daq}
The data acquisition (DAQ) system processes, digitizes, and records the raw signals from the PMTs. 
The DAQ chain begins with an analog pulse from the PMT, which is then digitized by digitizer boards. 
A dedicated trigger system to select desired events records the digitized waveform.
The PMTs are powered by multiple CAEN A7236N high-voltage (HV) modules, with connections via SHV cables. 
To minimize timing discrepancies between channels, all cables were matched in length, ensuring synchronization in signal arrival times.
To ensure long-term operational stability, the output current from each module is continuously monitored and logged using the GECO2020 software. 

The analog signal from each PMTs is transferred to the digitizer system. 
The waveform is converted from analog to digital using three CAEN V1730 14-bit, 500 MS/s digitizers and one V1740 12-bit 62.5 MS/s digitizer.
The acquisition time window is 1,920 ns.
This window was specifically set for accurate time synchronization between the two different digitizer models.
The V1730 digitizers sample at 500 MS/s, corresponding to a sample interval of 2 ns, while the V1740 samples at 62.5 MS/s, corresponding to a 16 ns interval.
The time window of 1,920 ns corresponds to a common multiple of their respective sampling intervals.
With this time window, both digitizers take an exact integer number of samples (960 for the V1730 and 120 for the V1740).
All digitizer boards are synchronized by daisy-chain connections.

The DAQ system uses following three types of trigger to select a specific type of event:
\begin{itemize}
 \item alpha source trigger: A $^{210}\mathrm{Pb}$ source was installed at the center of the detector.
The source is embedded in a plastic scintillator.
A small PMT is attached to the plastic scintillator.
The source emits alpha particle producing scintillation light that is detected by the small PMT, triggering DAQ.
The alpha source was used for PMT calibration and will be described in detail later in this paper.
 \item top paddle trigger: Two 10 cm $\times$ 30 cm plastic scintillator paddles, each read out by a PMT, are installed above the detector \textcolor{black}{visible at the very top of the assembly in Fig.~\ref{fig:1ton_schematic}.}, oriented orthogonally to create a 10 cm $\times$ 10 cm overlapping region centered above the tank.\textcolor{black}{\sout{as shown in Fig.~\ref{fig:1ton_schematic}.}}
 These top paddles aim to tag cosmic muon entering the tank.
 An identical pair of paddles is installed directly below the tank, arranged parallel to each other, providing an additional tag for crossing muon. 
 The clear trajectory of crossing muons has consistent energy deposition within the detector volume which allows us to measure the light yield.
\item majority trigger: It was implemented in October 2024 with the main aim of detecting various muon angle events including side-entering muon. 
The trigger is generated using the analog sum outputs of the two V1730 digitizers for the 30 bottom PMTs. 
A trigger decision is made based on the number of hit channels (nhit) exceeding a threshold on both boards simultaneously. 
Based on the analysis of crossing muons from the top paddle triggered events, the threshold is optimized to be 13 hits per board to ensure near 100\% trigger efficiency for the desired event topology.
\end{itemize}

The three trigger signals are combined using an OR logic board, which outputs a single signal whenever at least one of the input trigger signals is active.
The resulting output is fed into the master boards as an external trigger.
Upon receiving the initial external trigger, the master board generates a trigger-out signal, which is connected to the external trigger input of the second board. 
Similar chain continues to the next boards, forming a daisy-chain connection across the boards.
The entire control for DAQ system is managed by a ethernet connected PC, which is running software based on the ToolDAQ~\cite{ToolDAQ}.

\begin{figure}[htbp]
  \centering
  \includegraphics[width=\linewidth]{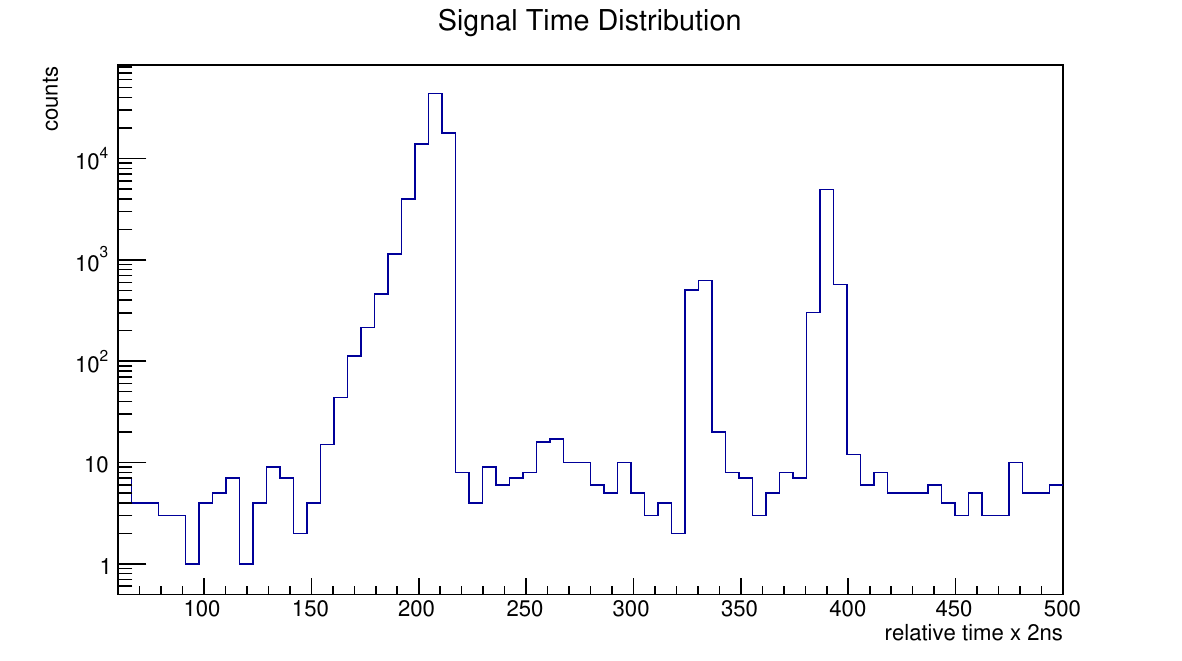}
  \caption{Triggered events time structure showing the first 1,000 ns. The absolute trigger time $t_0$ is at 480. The left peak is majority events. The middle is alpha events. The right is top paddle events. The time structure is primarily due to the contribution from the cable length differences for the alpha and top paddle signals and the intrinsic signal processing delay for the majority signal.}
  \label{fig:time_structure}
\end{figure}
The three triggers are clearly separated in time, as shown in Fig.~\ref{fig:time_structure}. 
This separation is by design, achieved through two mechanisms. 
First is the cable length difference between the alpha source PMT and the top paddle PMTs.
The alpha source PMT cable was intentionally made longer than the top paddle PMT cables.
This cable length difference introduces a signal arriving time delay, which enables us to distinguish alpha events from muon events based on the trigger time. 
Second, due to the additional instrument\textcolor{black}{s} such as oscilloscope and logic boards, the majority trigger signal incurs a processing delay compared to alpha and top paddle triggers. 
This time structure allows us to select desired events based on the event time structure.

\section{PMT Calibration}
\label{sec:calibration}
To ensure data reliability over long-term operation, the gain stability of each PMT was continuously monitored. 
PMT gain ($G$) is defined as the number of electrons collected at the anode for each photoelectron generated at the photocathode.
For the calibration, a $^{210}$Pb alpha source embedded in a EJ228 plastic scintillator was installed at the center of the detector~\cite{2022_WbLS_paper}.
It decays to $^{210}\mathrm{Po}$, which emits 5.304 MeV of alpha particles, producing scintillation light that is detected by each PMT.
The number of photons arriving at each PMT is typically at the single photoelectron (SPE) level, allowing the SPE peak to be clearly identified in the resulting charge distribution.

\begin{figure}[htbp]
    \centering
    \includegraphics[width=\linewidth]{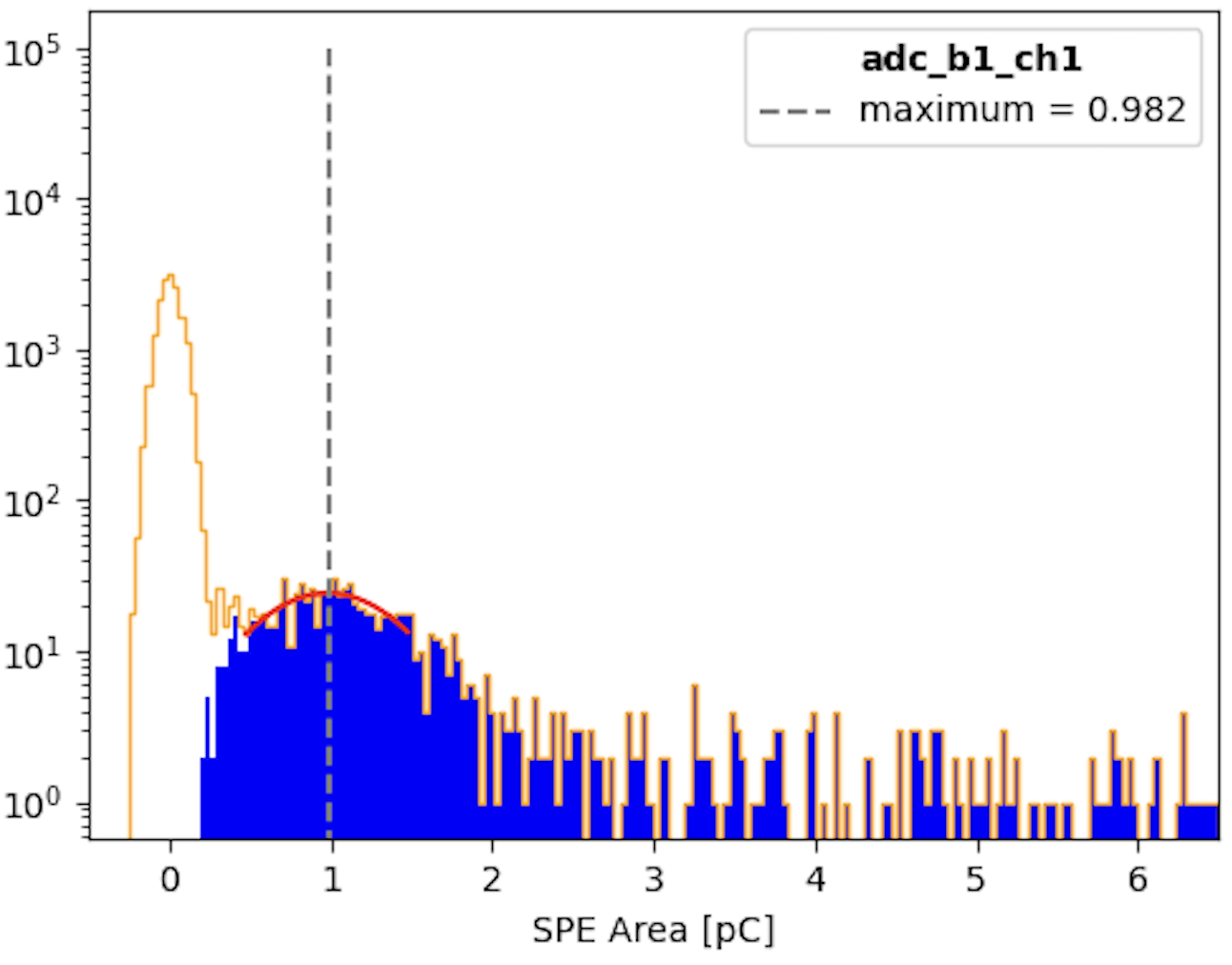}
    \caption{A typical charge spectrum of alpha events from a single PMT channel. The high-statistics peak near 0 is the pedestal, while the subsequent peak corresponds to the single photoelectron (SPE). The red curve shows a Gaussian fit near the SPE peak and the mean of a Gaussian fit is indicated by the vertical dashed line}
    \label{fig:spe_distribution}
\end{figure}

The charge distribution from alpha events exhibits a clear SPE peak, separated from the pedestal at 0 as shown in Fig.~\ref{fig:spe_distribution}.
The mean ADC count of the SPE peak is determined by a Gaussian fit.
The gain is calculated using following relation.
\begin{equation} 
\label{eq:ADC_to_gain}
G = \frac{\mathrm{ADC}_{mean} \times C_{\mathrm{ADC}}}{e}
\end{equation}
, where $\mathrm{ADC}_{mean}$ is the mean ADC count of the SPE peak, $C_{\mathrm{ADC}}$ (pC/ADC count)
represents the digitizer's charge-to-ADC conversion factor, and $e$ is the elementary charge.

Alpha events were taken continuously throughout the entire data-taking period, facilitated by the distinct timing structure described in Section~\ref{sec:daq}. 
The gain of each PMT is calculated and calibrated on a daily basis, providing a precise, continuous record of detector stability.

\section{WbLS Injection}
The entire data-taking period had two main parts: water phase and WbLS phase.
Water data were taken with the detector filled with pure water to one-quarter, one-half, and its full capacity as a baseline dataset. 
Following the water phase, the WbLS was formed by introducing organic components sequentially~\cite{2022_WbLS_paper}.
\begin{table}[h]
    \centering
    \begin{tabular}{|c|c|c|}
        \hline

        \multicolumn{2}{|c|}{Liquid Configuration} & \multirow{2}{*}{Run Start (yr-m-d)} \\
        \cline{1-2} 
        Type & Concentration/Fill Level & \\
        \hline

        \multirow{4}{*}{Water} & empty &  25-04-16 \\
         & quarter fill &  24-04-29 \\
         & half fill &  24-05-22\\
         & full fill &  24-06-15\\
        \hline

        \multirow{6}{*}{WbLS} & 0.35\% & 24-11-15 \\
         & 0.45\% & 24-12-03\\
         & 0.55\% & 24-12-19\\
         & 0.65\% & 25-01-07\\
         & 0.75\% & 25-01-22\\
         & 1.0\% & 25-03-11 \\
        \hline
    \end{tabular}
    \caption{Summary of run periods for each detector configuration.}
    \label{tab:run_plan}
\end{table}
Table.~\ref{tab:run_plan} summarizes entire data taking phases.
For each concentration level, data were taken \textcolor{black}{for approximately two weeks}, a period chosen to gather enough statistics, around 2,000 crossing muons. 
By taking data in this stepwise way, the scintillation light yield can be mapped out as a function of WbLS concentration.
The concentration is determined by mass, i.e. for 0.35\%, 3.5 kg of concentrated WbLS (cWbLS) is injected.
The cWbLS was injected by using an external pump integrated with the detector's circulation system. 
To achieve precise WbLS concentration, a specific amount of water was taken out first and then \textcolor{black}{a} corresponding amount of concentrated WbLS is injected.
To ensure the WbLS mixed uniformly during injection, it was injected slowly at constant rate. 
While the injection was in progress, the detected PE was monitored in real time to ensure that the injection process was stable.

\begin{figure}
    \centering
    \includegraphics[width=1.0\linewidth]{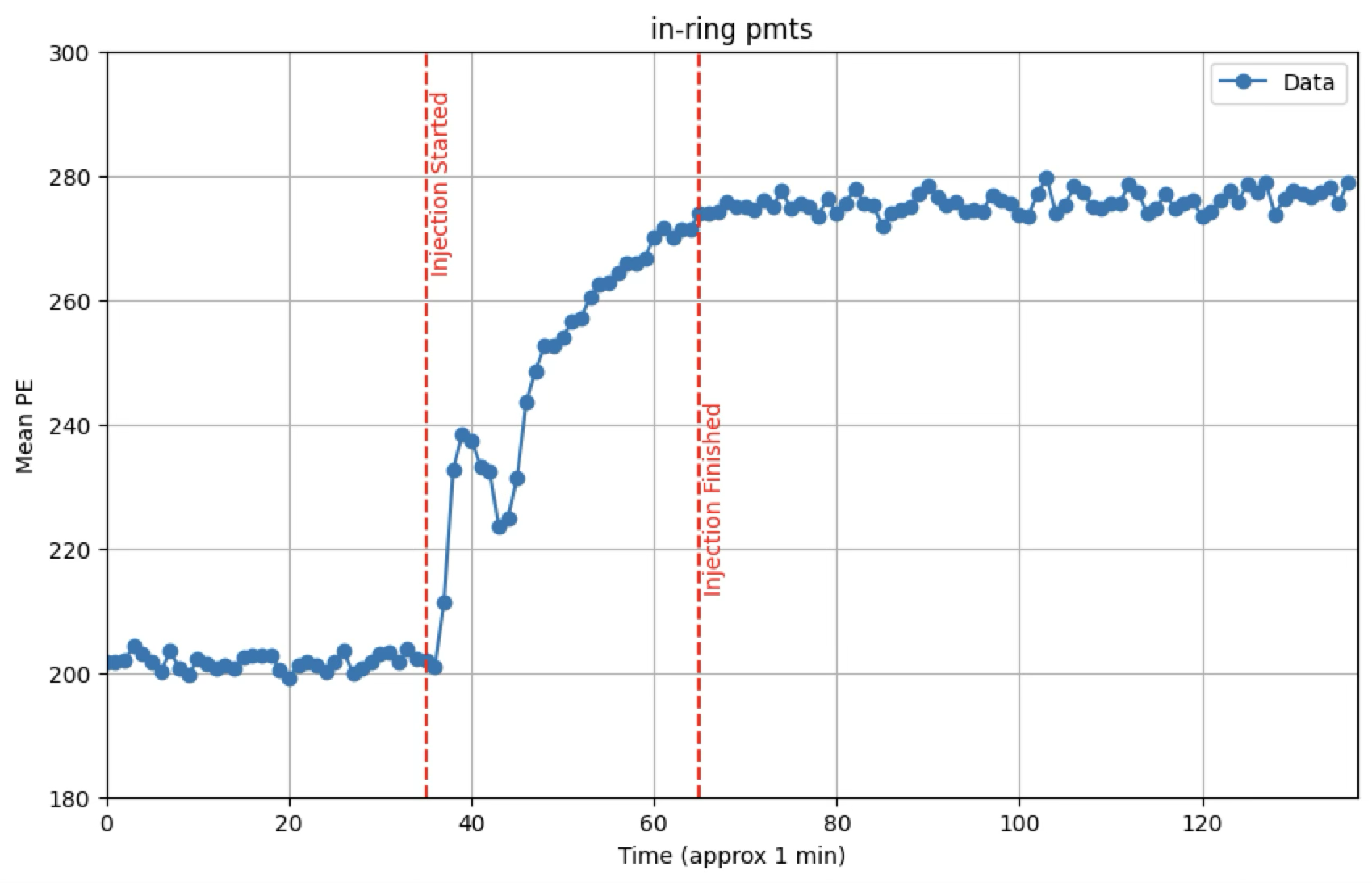}
    \caption{The mean PE for in-ring PMTs changing during the first injection. each points corresponds to approximately 1 minute. After the first significant light increase, a dip is observed, which due to non-uniformity in the WbLS. At the start of the injection, a localized region of relatively high concentration forms, leading to a sharp increase in PE. As this region gradually diffuses, the PE decreases. With continued injection, however, the PE increases once again.}
    \label{fig:first_injection}
\end{figure}

The first injected WbLS concentration was 0.35\% which is the minimum concentration for the mixture to be stable chemically.
Fig.~\ref{fig:first_injection} shows the mean PE detected during the first injection for events selected by the majority trigger from in-ring PMTs.
The majority trigger was necessary because crossing muons event rate is too low for real-time monitoring.
The majority trigger provides a much higher rate of approximately 150 Hz, enabling such monitoring.
As shown in the figure immediate increase in the detected PE is observed at the start of the injection. 
This is due to the formation of a localized high-concentrated region of WbLS near the injection region.
WbLS then started to diffuse, temporarily lowering the effective concentration seen by the PMTs and causing a temporary dip after the first increase.
As the injection continues and the WbLS becomes more uniform detected PE increased again.

\begin{figure*}[t]
    \centering
    \includegraphics[width=\linewidth]{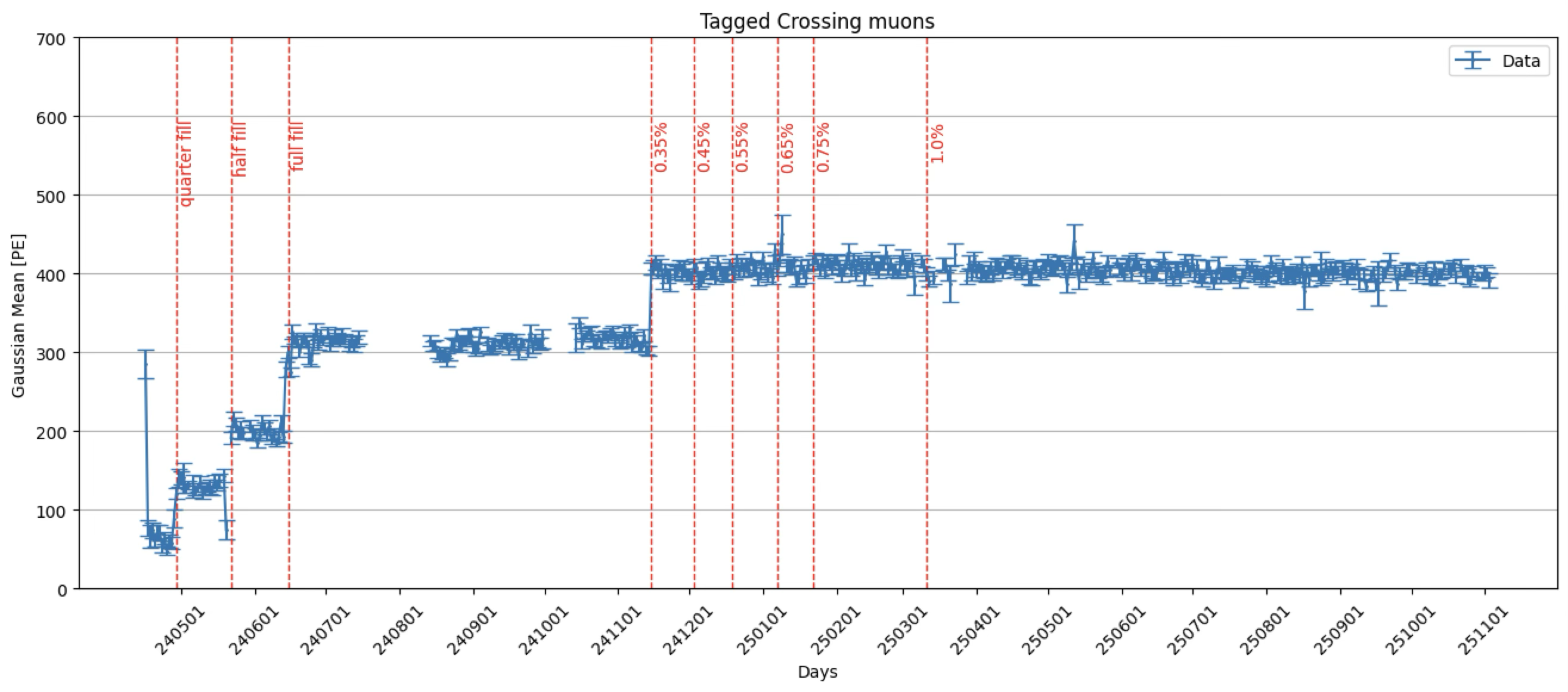}
    \includegraphics[width=\linewidth]{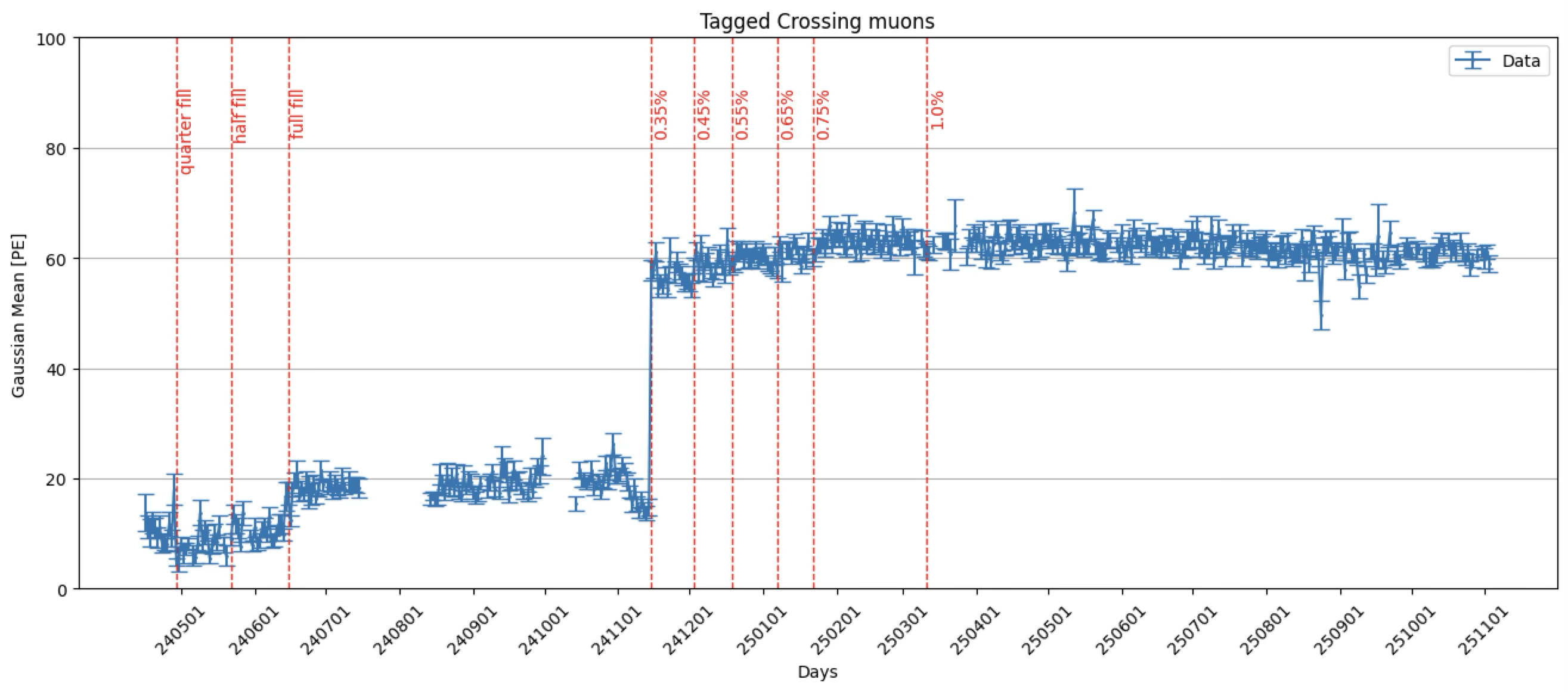}
    \caption{The daily mean of detected PE from crossing muon events for in-ring (top) and out-ring (bottom) PMTs over the entire data-taking period. Each data point represents the mean of a Gaussian fit to the daily PE distribution, with error bars indicating the error on the mean from the fit. Vertical dashed lines denote a change in the detector's liquid configuration, as detailed in Table~\ref{tab:run_plan}.}
    \label{fig:LY_curve}
\end{figure*}
Fig.~\ref{fig:LY_curve} shows the detected PE from selected crossing muon events throughout the data-taking period. 
In contrast to Fig.~\ref{fig:first_injection}, which uses majority-triggered events, Fig.~\ref{fig:LY_curve} uses data from crossing muon events.
This is because crossing muons exhibit less angular variation, providing a more robust measurement for long-term stability monitoring. 
The selected event rate is approximately 139.4 crossing muon events per day.
Each data point represents the mean of a Gaussian fit to the PE distribution measured on a specific day, with the error bars indicating the error on the Gaussian mean from the fit.
The experimental timeline is clearly visible, with red dashed lines marking each time the liquid in the detector was modified. 
The data taking begin with the water phase (before November 15, 2024) and the WbLS phase follows, where the concentration starting from an initial concentration of 0.35\% and increased by 0.1\% stepwise.
The data shown in the figure has two gaps. 
The first gap is one-month period beginning July 15, 2024, to replace the alpha source tube after we discovered cracks in it. 
We paused data-taking again for about a week in early October to allow for the implementation of a new majority trigger system.
\textcolor{black}{A temporary decrease observed before the first injection corresponds to a change in optical reflection conditions during WbLS injection preparation. 
To inject WbLS, the water in the detector was partially drained.
Before draining, the water level reached the top of the detector and was in direct contact with the acrylic surface. 
After partial drainage, a few-millimeter air gap formed between the water and acrylic, altering the reflection environment and resulting in the observed reduction.}

After the first injection, we observed significant PE increase both for in-ring and out-ring region.
This significant increase is attributed to the re-emission properties of WbLS, which can absorb Cherenkov photons with wavelengths below 350 nm and subsequently re-emit them at longer wavelengths.
This “Cherenkov conversion” process transforms previously undetectable photons into detectable ones. As the PMTs in our detector exhibit higher sensitivity in the 350–700 nm range, only a small fraction of Cherenkov photons were initially detected. Following the conversion, the re-emitted photons fall within the PMT sensitivity range, resulting in a substantial increase in the detected photoelectrons.
Following injections increased concentrations to: 0.45\%, 0.55\%, 0.65\%, 0.75\%, and finally 1.0\%. 
From 0.45\% to 1\%, PE increased gradually \textcolor{black}{as the concentration increased.}

The final injection was on March 11, 2025, increasing the concentration to 1.0\%. 
The long-term stability was evaluated with the 1.0\% concentration. 
The mean PE of 1.0\% WbLS shows the stability, with the daily fluctuations less than 2\% around the average as shown in Fig.~\ref{fig:LY_curve}.
To quantify the stability, a linear fit was performed to 1.0\% WbLS data. 
The resulting slope was (-0.0506 $\pm$ 0.0147) PE/day for in-ring and (-0.0076 $\pm$ 0.0031) PE/day for out-ring. 
This results confirms that 1.0\% WbLS is chemically and optically stable over \textcolor{black}{a} long-term period in the 1-ton scale detector.

\section{Benchtop Light Yield Measurement}
\label{sec:benchtop}
\begin{figure}[htbp]
    \centering
    \includegraphics[width=1.0\linewidth]{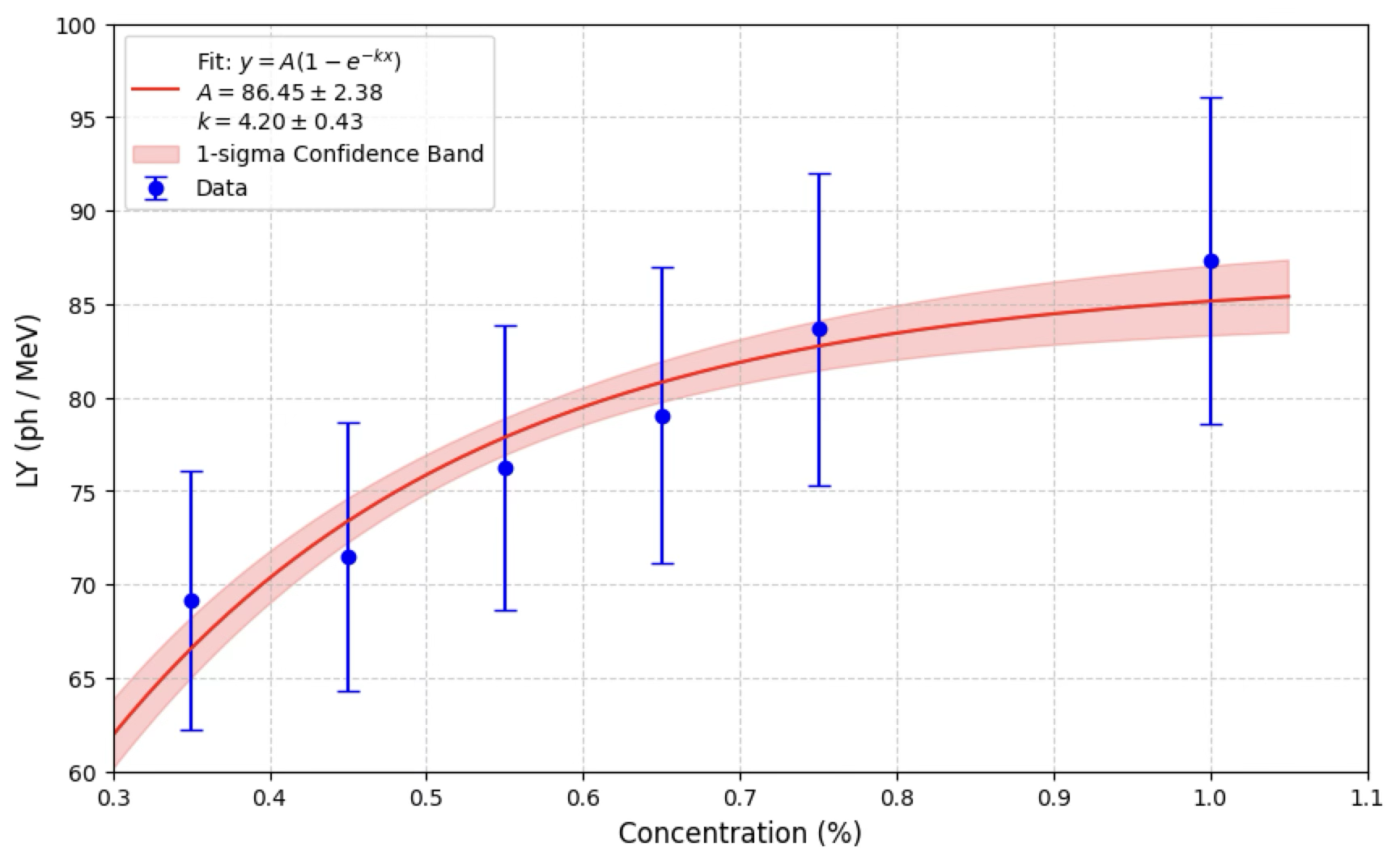}
    \caption{The light yield measured from benchtop measurement. $y=A \times (1-e^{-kx})$ is used to fit the points. These values are used for MC simulation in light yield analysis.}
    \label{fig:chemistry}
\end{figure}
To interpret the PE collected from scintillator only in the detector, Monte Carlo (MC) simulations are essential, and the absolute light yield (LY) of WbLS is the most critical input parameter.
To measure LY, a separate benchtop measurement was performed.
In this measurement, we utilize a Beckman Coulter LS6500 Liquid Scintillation Counter (Beckman Coulter, Inc., Brea, CA, USA) with a $^{137}$Cs radioactive source, which provides a source of high-energy electrons through Compton scattering from its 662 keV gamma-ray emissions. 
A set of small vials filled with a WbLS sample \textcolor{black}{were} exposed to the source. 
The two PMTs inside the LS6500 then detect the photons from Compton scattering in coincidence to effectively reduce noise. 
The resulting spectra were recorded for each sample to analyze their light yield.
To quantify light yield of each sample, we utilized Linear Alkylbenzene (LAB)-based liquid scintillator (LS) with a well-known light yield as a reference.
The spectrum of reference LS is measured under the same condition.
The absolute light yield can be calculated by comparing the Compton edge location of WbLS and LS:
\begin{equation}
    \text{LY}_{\text{WbLS}} = \frac{\text{Edge}_{\text{WbLS}}}{\text{Edge}_{\text{LS}}} \times \text{LY}_{\text{LS}}
\end{equation}
A ``power derivative'' method~\cite{power_method_paper} was used to determine the Compton edge. 
This method converts measured spectrum to power spectrum ($f(x) = x \times N(x)$), where $x$ represents the channel number (proportional to energy) and $N(x)$ is the number of counts in that channel.
This transformation makes the Compton peak more clear without change on the Compton edge.
The derivative minimum of converted power spectrum can be used as the Compton edge location.
The complete set of light yield measured by this method is summarized in Table~\ref{tab:table_LY_fitting_result}.
A simple saturation model, $y=A \times (1-e^{-kx})$, is fitted to the measured data points, which help us to expect the interpolation for the intermediate concentrations.
The best-fit values for $A$ and $k$ are 86.45 $\pm$ 2.38 and 4.20 $\pm$ 0.43 respectively.
Fig.~\ref{fig:chemistry} shows the data, the best-fit curve. 

\begin{table}[h]
    \centering
    \begin{tabular}{|c|c|c|}
        \hline
        WbLS Concentration & Light Yield\\
        \hline
        0.35\% & \textcolor{black}{(69.16 $\pm$ 6.92)} ph/MeV\\
        0.45\% & \textcolor{black}{(71.49 $\pm$ 7.15)} ph/MeV\\
        0.55\% & \textcolor{black}{(76.28 $\pm$ 7.63)} ph/MeV\\
        0.65\% & \textcolor{black}{(79.06 $\pm$ 7.91)} ph/MeV\\                
        0.75\% & \textcolor{black}{(83.68 $\pm$ 8.37)} ph/MeV\\
        1.0\% & \textcolor{black}{(87.32 $\pm$ 8.73)} ph/MeV\\
        \hline
    \end{tabular}
    \caption{Summary of the absolute light yield for each WbLS concentration.}
    \label{tab:table_LY_fitting_result}
\end{table}

\section{Light Yield Analysis}
The main goal of this analysis is to understand the optical properties of WbLS by separating Cherenkov and non-Cherenkov components.
The light detected by PMT can be categorized as two optical components: Cherenkov and non-Cherenkov light:
\begin{itemize}
    \item Cherenkov light: Cherenkov radiation directly detected by the PMT.
    \item Non-Cherenkov light consists of two main contributions:
    \begin{itemize}
        \item Intrinsic scintillation: the direct scintillation light produced by WbLS.
        \item Re-emission: both scintillation and Cherenkov photons can be absorbed by WbLS and re-emitted.
    \end{itemize}
\end{itemize}

In this analysis, crossing muon events are selected.
The crossing-muon allows us to reconstruct the energy deposition in the detector due to its relatively constant travel length in the detector. 
These crossing-muon events can be selected by the tags from both top and bottom paddles.
The muon angular variation is 8$^\circ$, which is due to the size of top and bottom paddles.
During the full-water phase from June 16, 2024, to November 15, 2024 (a total of 152 days), 21,184 crossing muon events were recorded, corresponding to an average rate of approximately 139.4 events per day.
This full-water data is crucial since it can serve as a benchmark for MC simulation.
\begin{figure}[htbp]
    \centering
    \includegraphics[width=1.0\linewidth]{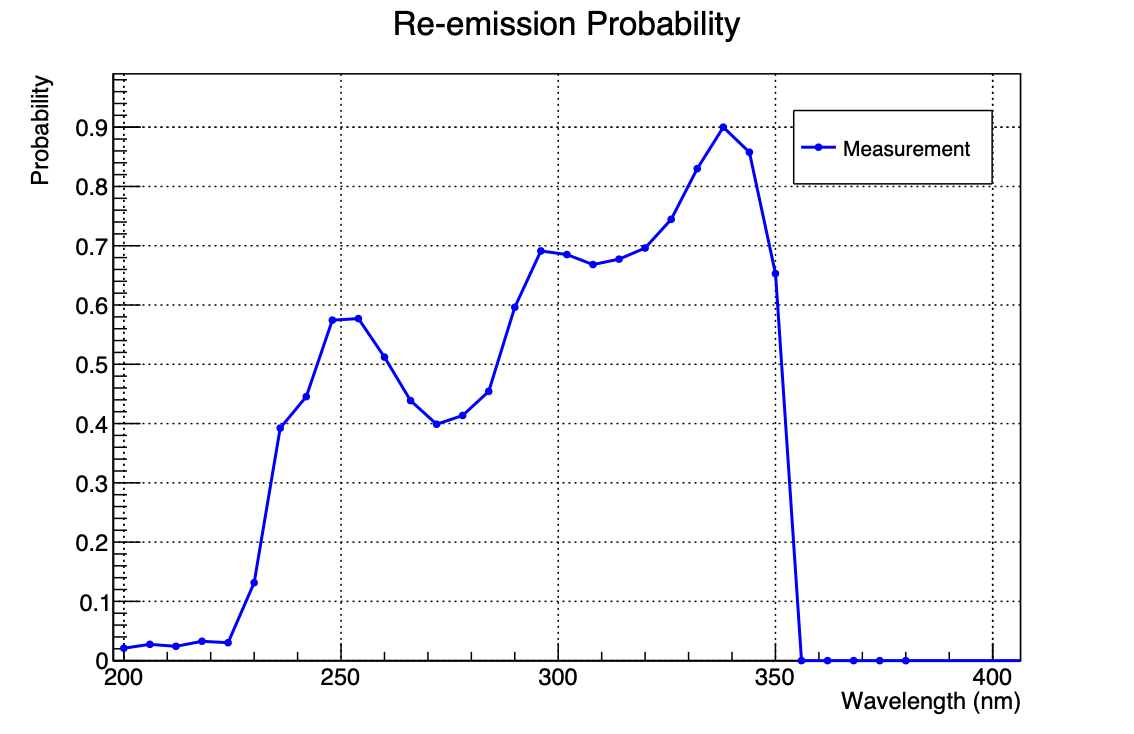}
    \caption{The reemission probability of WbLS as a function of the photon wavelength}
    \label{fig:new_reemission_prob}
\end{figure}
MC simulation used the GEANT4-based RATPAC-TWO~\cite{ratpac} framework with the realistic 1-ton detector geometry described in Section~\ref{sec:1-ton_detector}.
The optical properties of the WbLS such as absorption, scattering, and intrinsic scintillation light yield were adopted from external measurement. 
The benchtop measurements of the intrinsic scintillation light yield described in Section~\ref{sec:benchtop} were implemented as the input light yield for simulation.
The re-emission probability as a function of the photon wavelength is shown in Fig.~\ref{fig:new_reemission_prob} \textcolor{black}{(ANNIE Collaboration, private communication, 2025)}.
To simulate realistic crossing muons, the initial position and momentum of the muons were configured to match the distribution determined from the trigger information of the top paddle detector in the actual experimental data. 
The energy and angular distribution is generated as flat distribution and then reweighted according to CRY (Cosmic-ray Library) generator~\cite{CRY}.
This approach was taken to minimize the discrepancy between the experiment and the simulation.

\subsection{MC Simulation Tuning}
\label{sec:MC Simulation Tuning}

\begin{figure}[htbp]
    \centering
    \includegraphics[width=1\linewidth]{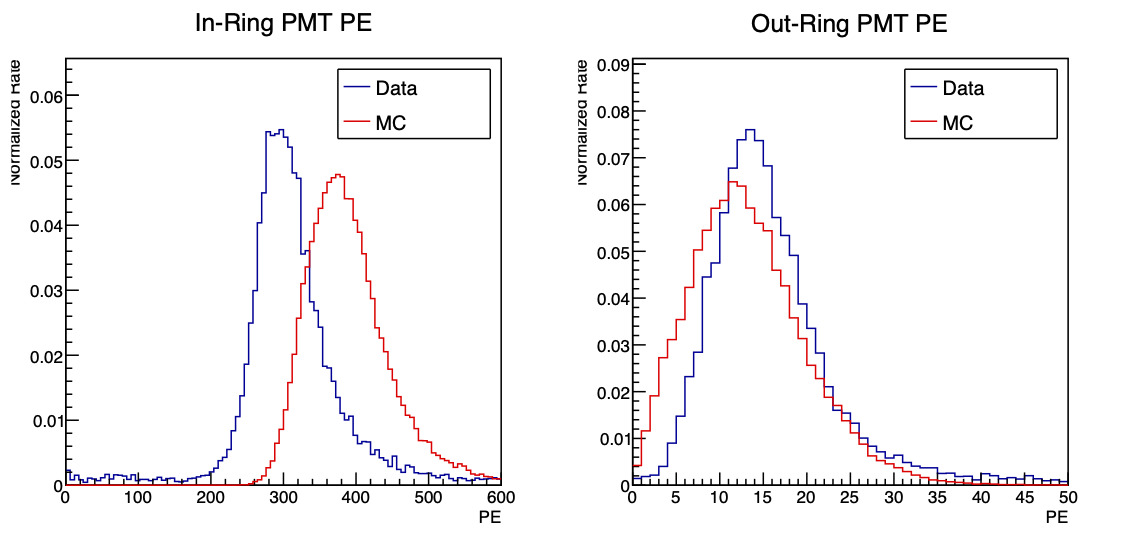}
    \caption{The comparison between data and MC simulation of crossing muon events for the in-ring (left) and out-ring (right) regions shows a discrepancy due to the PMT inefficiency. To resolve this, a correction factor is applied to the simulation, defined as the ratio of the Gaussian means fits to the data and MC PE distributions.}
    \label{fig:pe_hist_rings}
\end{figure}
\begin{figure}[htbp]
    \centering
    \includegraphics[width=1.0\linewidth]{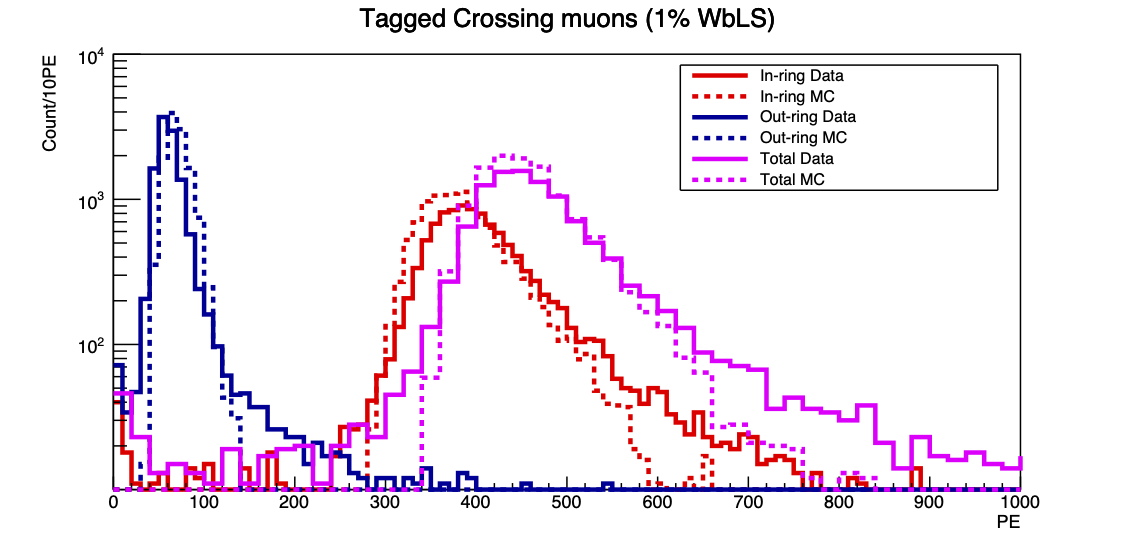}
    \caption{PE distribution of crossing muons with 1\% WbLS for data and MC simulation. In-ring and out-ring efficiency correction is applied to MC.}
    \label{fig:pe_dist}
\end{figure}

Since \textcolor{black}{the} MC model for WbLS is not perfect, we need to tune the model with well-understood water data.
Fig.~\ref{fig:pe_hist_rings} shows a comparison between the water data and the MC simulation for crossing muon events, shown separately for the in-ring and out-ring regions.
Since water is simple and well understood \textcolor{black}{and assuming the MC model for Cherenkov production and propagation is well-modeled}, the discrepancy between data and MC is primarily due to the PMT \textcolor{black}{inefficiency}.
The discrepancy from the PMT inefficiency, such as poor optical coupling between the PMT and the arcylic surface and reduced PMT performance on quantum efficiency, is corrected by applying a correction factor to the PE distribution from MC simulation. 
The correction factor is defined as the ratio of the Gaussian mean of the PE distributions for data and MC.
The correction accounts for overall PMT efficiency such as the quantum efficiency and optical coupling between the tank and the PMT.
In the analysis, corrections for both in-ring and out-ring were applied to all data.
The correction factors for in-ring and out-ring PMTs are measured as 0.79 and 1.18 repectively.
In addition, alpha triggered events can serve the efficiency correction as well by the same data-MC comparison method since PMT efficiency is not dependent on \textcolor{black}{a} specific particle.
The correction factors from alpha data-MC comparison are measured as 0.72 (in-ring) and 1.06 (out-ring).
The differences between water and alpha correction factors are used as a systematic uncertainty in the main analysis.
Fig.~\ref{fig:pe_dist} shows 1\% WbLS data-MC comparison after the corrections are applied.

\subsection{Light Yield Analysis Methods and Results}
\begin{figure}[htbp]
    \centering
    \includegraphics[width=1.0\linewidth]{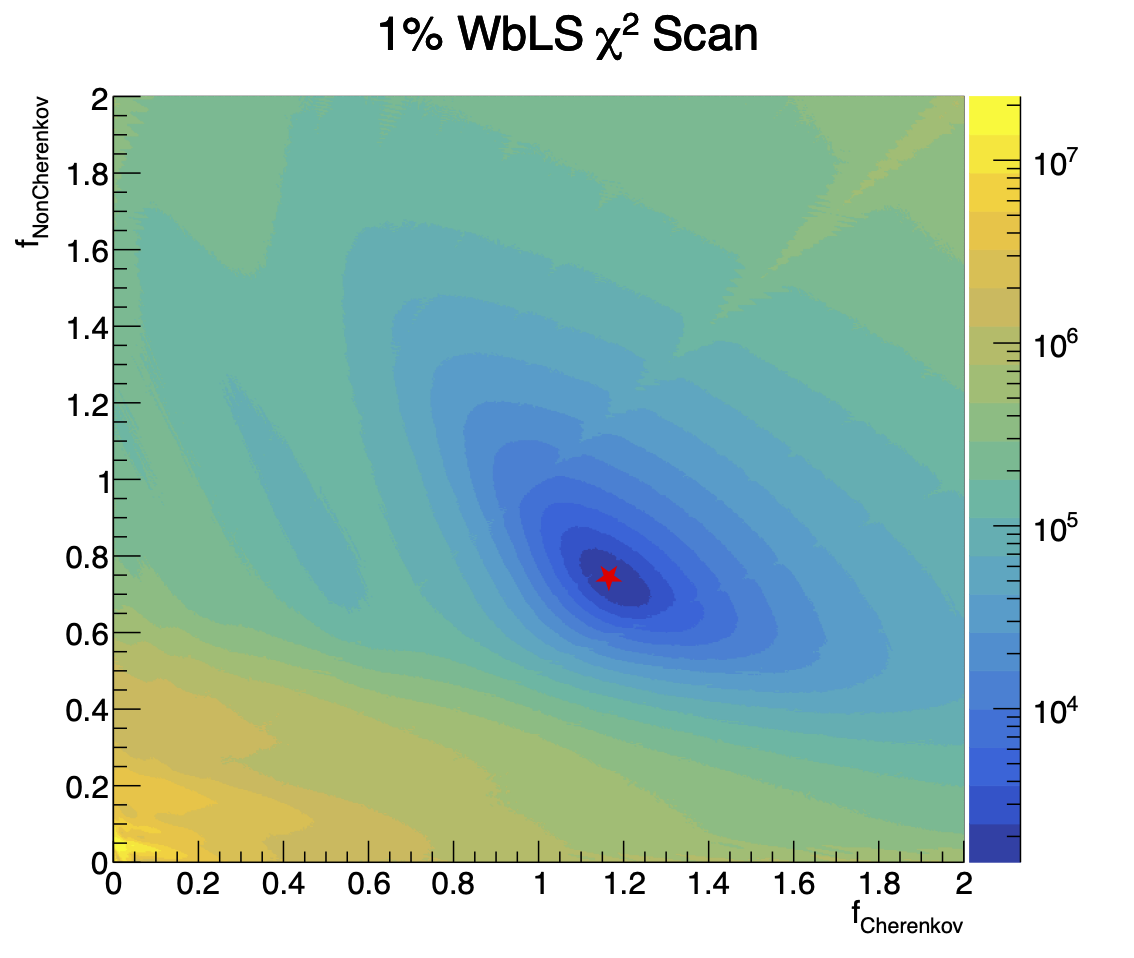}
    \caption{A two-dimensional scan of the $\chi^2$ as a function of the Cherenkov ($f_{\text{Cherenkov}}$) and non-Cherenkov ($f_{\text{NonCherenkov}}$) light scaling factors for the 1.0\% WbLS dataset. The color indicates the $\chi^2$ value on a logarithmic scale. The red marker indicates the best-fit values corresponding to the minimum $\chi^2$.}
    \label{fig:Chi2Map}
\end{figure}

To quantitatively evaluate the agreement between MC simulation and data, a fitting procedure was employed. 
The primary objective of this fit is to tune the simulated light components to best match the observed PE distribution from the detector.
The fitting is based on a linear scaling model that adjusts the contributions of Cherenkov and non-Cherenkov light from the simulation. The predicted total charge is given by the equation:
\begin{equation}
\begin{split}
    Q_{\text{predicted}} = & f_{\text{Cherenkov}} \times Q^{\text{MC}}_{\text{Cherenkov}} \\
    & + f_{\text{NonCherenkov}} \times Q^{\text{MC}}_{\text{NonCherenkov}}
\end{split}
\end{equation}
where $Q^{\text{MC}}_{\text{Cherenkov}}$ and $Q^{\text{MC}}_{\text{NonCherenkov}}$ are the simulated charge contributions from Cherenkov and non-Cherenkov light, respectively. The parameters $f_{\text{Cherenkov}}$ and $f_{\text{NonCherenkov}}$ are the scaling factors for each light component that are optimized by the fit.
To find the best-fit values for $f_{\text{Cherenkov}}$ and $f_{\text{NonCherenkov}}$, we performed a chi-squared ($\chi^2$) minimization using a grid scan method.
This approach involves constructing a two-dimensional grid of possible scaling factor pairs.
For each pair on the grid, we calculate a $\chi^2$ value by comparing the predicted PE distribution from the scaled MC with the PE distribution measured in the data. 
The $\chi^2$ is defined as:
\begin{equation}
    \chi^2 = \sum_i \frac{(N^{\text{data}}_i - N^{\text{predicted}}_i)^2}{(\sigma^{\text{data}}_i)^2}
\end{equation}
where the sum is over all bins i of the PE distribution, $N^{\text{data}}_i$ is the number of events in the data, $N^{\text{predicted}}_i$ is the predicted number of events from the scaled MC, and $\sigma^{\text{data}}_i$ is the statistical uncertainty on the data count. 
The pair of factors that yields the minimum $\chi^2$ is chosen as the best-fit result.
The grid was defined with 600 steps for each parameter, spanning a range from 0.0 to 2.0.
For a shape-only comparison, both the data and MC histograms were normalized to the same area.
Fig.~\ref{fig:Chi2Map} shows the 2-D $\chi^2$ map for 1\% WbLS.

\begin{table}[h]
    \centering
    \begin{tabular}{|c|c|c|}
        \hline
        WbLS Concentration & $f_{\text{Cherenkov}}$ & $f_{\text{NonCherenkov}}$ \\
        \hline
        0.35\% & 1.1653 $\pm$ 0.1102 & 0.6544 $\pm$ 0.0701 \\
        0.45\% & 1.1720 $\pm$ 0.1101 & 0.6745 $\pm$ 0.0768  \\
        0.55\% & 1.1853 $\pm$ 0.1069 & 0.6711 $\pm$ 0.0935 \\
        0.65\% & 1.1853 $\pm$ 0.0935 & 0.6978 $\pm$ 0.0802 \\                
        0.75\% & 1.1720 $\pm$ 0.1068 & 0.6945$\pm$ 0.0843 \\
        1.0\% & 1.1619 $\pm$ 0.1269 & 0.7379 $\pm$ 0.0835 \\
        \hline
    \end{tabular}
    \caption{Summary of the optimal scaling factors for the Cherenkov ($f_{\text{Cherenkov}}$) and non-Cherenkov ($f_{\text{NonCherenkov}}$) light components. These values were determined by applying the fitting procedure, described in the text, to the dataset for each WbLS concentration.}
    \label{tab:table_fitting_result}
\end{table}
The fitting procedure was applied independently to the dataset for each WbLS concentration to determine the optimal scaling factors. 
The results of these fits are summarized in Table~\ref{tab:table_fitting_result}.
The scaling factor for the Cherenkov component, $f_{\text{Cherenkov}}$, remains relatively constant across all concentrations. 
This suggests that the MC modeling of Cherenkov photon production and propagation is largely independent of the WbLS concentration, as expected. 
The value being consistently above unity indicates a slight underestimation of the Cherenkov light in the MC simulation.
In contrast, the non-Cherenkov scaling factor, $f_{\text{NonCherenkov}}$,  was observed to be higher at 1.0\% WbLS concentration (0.7379 $\pm$ 0.0835) compared to at 0.35\% (0.6544 $\pm$ 0.0701).

A value of $f_{\text{NonCherenkov}}$ smaller than 1 indicates that our MC simulation overestimates the non-Cherenkov light component.
Since the intrinsic light yield is measured in precision by benchtop measurement,
this may be attributed to an overestimation of re-emission property within the MC model. 
The fact that $f_{\text{NonCherenkov}}$ increased to 1 at higher concentrations implies that the degree of overestimation by the MC is reduced. 
This suggests re-emission becomes more efficient as the WbLS concentration increases. 
Further detailed studies are needed to fully understand these observed correlation.

\subsection{Quantitative Determination of the Non-Cherenkov Component}
As an independent complementary analysis to the 2D fitting procedure described above, a simple linear model was applied to the out-ring PMTs to quantitatively estimate the non-Cherenkov light component.
The mean PE observed in the out-ring region can be modeled by the following linear model:
\begin{equation}
    \text{PE}_{\text{out-ring}} = y \times (\text{LY} + x) + a,
\end{equation}
where LY is the intrinsic scintillation light yield for a given WbLS concentration, as measured independently by the benchtop measurement described in Section~\ref{sec:benchtop}, $y$ is an overall efficiency and light collection factor, $x$ is an effective term representing non-Cherenkov light contributions other than intrinsic scintillation, and $a$ is Cherenkov related term such as Cherenkov light leakage or reflection into out-ring region.
It is geometry-dependent and assumed to be consistent across all WbLS datasets.
We performed a fit of the linear model to the mean out-ring PE values from the experimental data across all WbLS concentrations shown in Fig.~\ref{fig:LY_curve}. 
The fit yielded a best-fit value for the non-Cherenkov term \textcolor{black}{$x = (12.05 \pm 10.02)$} ph/MeV with $\chi^2/\text{ndf} = 0.68$. The excellent quality of the fit demonstrates that this simple model accurately describes the non-Cherenkov light response over the full range of WbLS concentrations studied.
\begin{table}[h]
    \centering
    \begin{tabular}{|c|c|c|c|}
        \hline
        WbLS Concentration & Data (PE) & Prediction (PE) & \% Error\\
        \hline
        0.35\% & 52.07 & 52.75 & 1.31\%\\
        0.45\% & 54.92 & 53.89 & 1.87\%\\
        0.55\% & 56.21 & 56.26 & 0.09\% \\
        0.65\% & 58.32 & 57.63 & 1.18\% \\                
        0.75\% & 59.02 & 59.91 & 1.50\% \\
        1.0\% & 61.80 & 61.70 & 0.16\% \\
        \hline
    \end{tabular}
    \caption{Comparison between the measured mean PE in the out-ring region and the predicted values from the best-fit model for each WbLS concentration. The final column shows the percentage error between the data and the prediction.}
    \label{tab:linear_regression}
\end{table}
The validity of our model was tested by comparing its predictions against the data for each WbLS concentration, using the best-fit parameters derived from the fit. Table~\ref{tab:linear_regression} details this comparison, listing the measured mean PE, the predicted mean PE, and the percentage error between them.
Excellent agreement is observed across the full range of WbLS concentrations. The model's predictions match the experimental data to within 1\% for all concentration. 
This strong correspondence confirms that our linear model accurately parameterizes the non-Cherenkov light contribution as a function of the intrinsic scintillation light yield, providing a robust description of the detector's response.

\subsection{Systematic Uncertainties}
The main systematic uncertainty in this analysis stems from the PMT efficiency~\cite{2022_WbLS_paper}. 
The 1 $\sigma$ variation on the PMT efficiency is quantified by taking the differences between correction factors obtained from water and alpha events described in Section~\ref{sec:MC Simulation Tuning}.
The resulting uncertainties on the non-Cherenkov component from the 1 $\sigma$ variation are approximately 10\% across the concentration.

\begin{figure}
    \centering
    \includegraphics[width=1.0\linewidth]{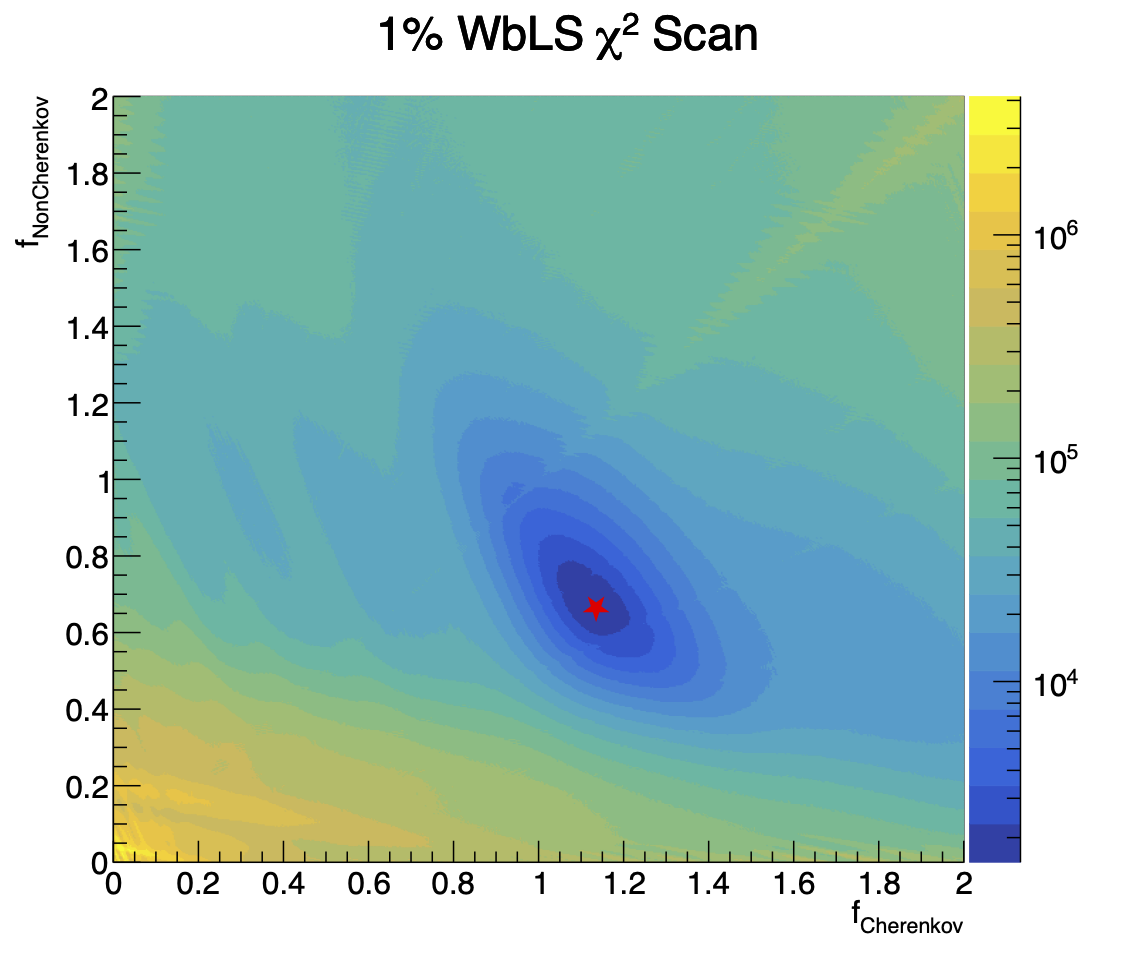}
    \caption{A two-dimensional scan of the $\chi^2$ using muon dataset from the hodoscope selection.}
    \label{fig:c_Chi2MapL100_hodoscope}
\end{figure}

To examine the impact of muon angular variation, 2D fitting results for crossing muons selected by the hodoscope were compared with those selected by the top and bottom paddles.
While the muon selected by top and bottom paddles can have angular variation up to 8$^\circ$, those selected by the hodoscope travel almost vertically downward.
Fig.~\ref{fig:c_Chi2MapL100_hodoscope} shows the 2D fitting result of hodoscope selected muon with 1\% WbLS.
The best-fit $f_{\text{NonCherenkov}}$	value for the hodoscope selection is 0.6644, which is a 10.7\% deviation from the value of 0.7446 obtained with the top and bottom paddle selection.
This indicates the angular variation can affect the result up to 10.7\%.

\section{Conclusion}
In this study, we utilized the 1-ton scale Water-based Liquid Scintillator detector in Brookhaven National Laboratory to measure the light yield of Gd-compatible WbLS as a function of its concentration and validate its long-term stability. 
We characterized the light yield and stability of Gd-compatible WbLS for concentrations ranging from 0.35\% to 1.0\%.
A key finding of this work is the confirmation of the critical role of the Cherenkov conversion property of WbLS. 
The first WbLS injection to 0.35\% resulted in a drastic increase in detected photoelectrons, which is from the conversion of otherwise undetectable UV Cherenkov photons into the longer wavelength within the PMT sensitive range.
For 1.0\% WbLS, the intrinsic scintillation light yield is measured as 87.32 $\pm$ 8.73 ph/MeV and the re-emission light yield as \textcolor{black}{12.05 $\pm$ 10.02} ph/MeV \textcolor{black}{by combining the 1-ton and the table-top experiments.}
The non-Cherenkov scaling factor $f_{\text{NonCherenkov}}$, \textcolor{black}{with the 1-ton detector measurement compared to the default simulation model,}  was observed to increase with WbLS concentration, from \textcolor{black}{0.6544 $\pm$ 0.0701} at 0.35\% to \textcolor{black}{0.7379 $\pm$ 0.0835} at 1.0\%. 

Furthermore, over 6 months operation at a 1.0\% Gd-compatible WbLS concentration demonstrated excellent stability. 
The daily mean PE over the entire period showed a variation of less than 2\% around the average, with a fitted slope -0.0506 $\pm$ 0.0147 PE/day for in-ring and -0.0076 $\pm$ 0.0031 PE/day for out-ring, which provides no strong evidence for a time dependence for both in-ring and out-ring PMTs. 
This result confirms the long-term stability of the Gd-compatible WbLS mixture.
These results are a significant step towards the realization of next-generation, large-scale and cost-effective neutrino experiments.

\section*{Acknowledgements}
This research was supported by Basic Science Research Program through the National Research Foundation of Korea(NRF) funded by the Ministry of Education (NRF-004065831G0002459).
The authors acknowledge support from the grant NRF-2022R1A2C1009686.
The work conducted at Brookhaven National Laboratory was supported by the U.S. Department
of Energy under contract DE-AC02-98CH10886. The work conducted at Lawrence Berkeley National
Laboratory was performed under the auspices of the U.S. Department of Energy under Contract DE-
AC02-05CH11231. The project was funded by the U.S. Department of Energy, National Nuclear
Security Administration, Office of Defense Nuclear Nonproliferation Research and Development
(DNN R\&D). This material is also based upon work supported by the U.S. Department of Energy, Office
of Science, Office of High Energy Physics, under Award Numbers DE-SC0018974, DE-SC0012704
and DE-SC0012447.
\newpage

\bibliography{reference} 

\end{document}